\documentclass[epsf,prb,twocolumn,showpacs,preprintnumbers]{revtex4}
\usepackage{graphics}
\usepackage{graphicx}
\usepackage{dcolumn} 
\usepackage{bm}
\usepackage{color}
\usepackage{epsfig}
\newcommand{\scoo}{Sr$_2$CrOsO$_6$}
\newcommand{\scro}{Sr$_2$CrRuO$_6$}
\begin{document}
\title{Half Semimetallic Antiferromagnetism in the 
Sr$_2$Cr${\cal T}$O$_6$ System, ${\cal T}$ = Os, Ru} 
\author{K.-W. Lee$^{1,2}$ and W. E. Pickett$^1$} 
\affiliation{$^1$ Department of Physics, University of California, Davis, 
  CA 95616, USA \\
 {$^2$ Department of Display and Semiconductor Physics, Korea University,
 Jochiwon, Chungnam-do 339-700, Korea}}
\date{\today}
\begin{abstract}
Double perovskite \scoo~is (or is very close to) a realization of a
spin-asymmetric semimetallic compensated ferrimagnet, according to
first principles calculations.
This type of near-half metallic antiferromagnet is an unusual occurrence, 
and more so
in this compound because
the zero gap is accidental rather than being symmetry determined.
The large spin-orbit coupling (SOC) of osmium upsets the spin balance
(no net spin moment without SOC): it reduces the Os spin moment by 0.27 $\mu_B$
and induces an Os orbital moment of 0.17 $\mu_B$ in the opposite direction.
The effects combine (with small oxygen contributions) to give a 
net total moment of 0.54 $\mu_B$ per cell
in \scoo, reflecting a large impact of SOC in this compound.  This value is
in moderately good agreement with the measured saturation moment 
of 0.75 $\mu_B$.  The value of the net moment on the Os ion obtained from
neutron diffraction (0.73 $\mu_B$ at low temperature) differs
from the calculated value (1.14 $\mu_B$).
Rather surprisingly, in isovalent \scro~ the smaller SOC-induced spin changes
and 
orbital moments (mostly on Ru) almost exactly cancel.
This makes \scro~a ``half (semi)metallic antiferromagnet'' (practically
vanishing net total moment) even when SOC
is included, with the metallic
channel being a small-band-overlap semimetal.
Fixed spin moment (FSM) calculations are presented for each compound, illustrating 
how they provide different information than in the case of a nonmagnetic
material.  These FSM results indicate that the Cr moment is an order of magnitude
stiffer against longitudinal fluctuations than is the Os moment.
\end{abstract}
\maketitle

\section{Introduction}
Precise compensation of magnetic moments in condensed matter systems,
resulting in a vanishing total moment, occurs only in rare instances
when not required by symmetry.
One example is spin (S) and orbital (L) moment cancellation\cite{adachi} in the 
Sm$^{3+}$ (or any $4f^5$)
ion, where Hund's rule requires antialigned
orientation of the spin and orbital moments (M$^S_z$=5$\mu_B$=-M$^L_z$)
along the quantization axis.  Another example is so-called
half metallic antiferromagnetism (HMAF), in which the integer spin moment
in a half metal has the value zero.\cite{leuken}  Although not an AF (up and down
spin directions are distinct), the up and 
down spin moments cancel exactly due to the half metallic nature. We
will use the more precise term {\it compensated half metal} (CHM).\cite{CHM} 
Compensation in this case is no longer exact if spin-orbit
coupling is not negligible.

After the initial description of the CHM concept,\cite{leuken} 
a rational search
for candidates was begun\cite{wep} in the double perovskite (DP) family of
oxides.  DPs have been around for decades\cite{poeppel} and new
ones appear regularly.  The suggestion that La$_2$MnVO$_6$ is a
promising candidate has been pursued by
Androulakis {\it et al.},\cite{andro}
whose samples did not seem to assume the desired charge states but which
also were not well ordered.  The (full and half) Heusler structures have
produced a number of HM ferromagnets,\cite{PT} 
which \textcolor{red}{have} encouraged searches for a
vanishing moment member.   Felser and collaborators identified a band-filling
criterion for CHMs in Heusler materials, and were led to
consider\cite{CHM,balke} Mn$_3$Ga,
which is calculated to be very near being a CHM (there is slight band
overlap).
In this compound two Mn $S=1$ moments
compensate one Mn $S=2$ moment.  Recently in the DP class, Park and
Min\cite{park} have calculated possible CHMs in the 
quintinary La${\cal A}$VMoO$_6$ system,
${\cal A}$= Ca, Sr, Ba, which are as yet unsynthesized.  Nakao has extended
the search for CHMs to tetrahedrally coordinated transition metal
based chalcopyrites.\cite{nakao}  A true CHM is not yet a laboratory
reality.

Originally reported 35 years ago
by Sleight and coworkers,\cite{sleight}
the compound \scoo~(SCOO) has been revisited by Kronkenberger
{\it et al.}\cite{krock} 
The motivation was that this
member follows a series of (known or suspected) half metals
Sr$_2$Cr${\cal T}$O$_6$, ${\cal T}$=W, Re, with very high magnetic ordering
temperatures (T$_N\sim$500 K and 635 K, respectively).  SCOO 
has an even higher ordering temperature T$_N$=725 K, the highest known in this
class, and was reported to be insulating.   
The Cr$^{3+}$ and Os$^{5+}$ ions both have a filled majority
$t_{2g}$ subshell and high spin $S=3/2$, eliminating possible complications
due to orbital fluctuations and orbital ordering and making it straightforward
to address the electronic and magnetic structure.  The calculated
densities of states (DOS) indicate\cite{krock} that, before 
Spin-orbit coupling (SOC) is
included, the system is  compensated ferrimagnet with a very small gap.
The structure is cubic above $\sim 500$ K , but distorts slightly 
below that temperature to rhombohedral $R\bar{3}$ symmetry, 
consistent with the magnetic symmetry.

In this paper we look in more detail at the electronic structure of
SCOO, which also prompts us to check the isovalent material \scro~(SCRO).
First principles calculations, of the same sort that predict Sr$_2$CrWO$_6$
and Sr$_2$CrReO$_6$ to be HMs, predict that SCOO is (before
considering SOC) actually a ferrimagnetic semimetal with 
precisely compensating
spin moments, or {\it  spin-asymmetric
compensated semimetallic ferrimagnet}
in which the electrons and the holes are each fully polarized and
have {\it opposite} spin directions, in spite of zero net moment and
hence no macroscopic magnetic field. 
This is a peculiar state indeed.
SOC degrades this by giving a nonzero total moment, but
the band structure is little changed.  We then look at the isovalent
system SCRO, and find that due to small but important chemical
differences, it is also a {\it compensated half (semi)metal} 
and moreover the total (spin + orbital) moment remains vanishingly small.
This is a CHM in which the metallic channel is semimetallic rather
than  fully metallic, with the semimetallic channel having spin in the
direction of the Cr spin (the Ru spin is antialigned).

\section{Theoretical and Computational Methods}
Our calculations are based on both cubic (500 K)
and rhombohedral (R$\bar{3}$, at 2 K) structures.\cite{krock}
In the distorted structure, the change in volume is less than 1\%, and
O and Sr atoms are displaced by 0.015 and 0.006~\AA, respectively.
Since comparison between perovskite SrOsO$_3$\cite{sroso3} 
and SrRuO$_3$\cite{srruo3} shows about 1\% volume difference, 
difference in structures between SCRO and SCOO is
expected to be negligible. Calculations (as described below) are consistent, 
obtaining a 1\% smaller lattice constant for \scro~(7.62 \AA) than 
for \scoo~(7.706 \AA).  Since we wish here to interpret the experimental
electronic properties as closely as possible, we use for our calculations
the same experimental structures of SCOO (also used for SCRO).
These are the values $a$=7.8243 \AA~ for the cubic,
and $a$=5.5176 \AA, $c$=13.445 \AA~ for the rhombohedral 
structures.\cite{krock}

These calculations were carried out with the local spin density approximation
(LSDA) and spin-orbital coupling (SOC) or fully relativistic scheme\cite{fplo2}
implemented in two all-electron full-potential codes,
FPLO\cite{fplo1} and Wien2k\cite{wien}, which showed consistent
results.
In FPLO, basis orbitals were chosen such as
Os $(4f5s5p)6s6p5d$, Ru $(4s4p)5s5p4d$, Cr $(3s3p)4s4p3d$,
Sr $(4s4p)5s5p4d$, and O $2s2p3d$.
(The orbitals in parentheses indicate semicore orbitals.)
In Wien2k, the basis size was determined by
$R_{mt}K_{max}$=7.0 and APW sphere radii (1.95 for Os/Ru, 1.94 for Cr,
2.47 for Sr, and 1.72 for O). Although the semimetallic character does not
require a lot $k$-points like a metal would, the Brillouin zone was sampled 
carefully with 
(20,20,20) $k$-mesh. A carefully $k$-mesh sampling is necessary for SOC
and fixed spin moment (FSM) calculations.
The FSM calculations we describe were done with the FPLO code.

For SCOO we find the R$\bar{3}$ distortion is energetically
favored, in agreement with observation.  We also find that SCRO
in the R$\bar{3}$ structure is favored over the cubic structure, by
145 meV.  The distortions (apparently due to size mismatch) in either compound
have negligible affect on the electronic structure, consistent with the
fact that both ions have a filled $t_{2g}$ shell so that band
structure energy is not gained by the distortion.  

\section{Results and Interpretation}
\subsection{LSDA electronic structure of SCOO}
\begin{table*}[bt]
\caption{LSDA and LSDA+SOC results for individual and total spin $M_S$, 
orbital $M_L$, 
and net $M_{net}$ moments ($\mu_B$) in the R$\bar{3}$ structure.
The effect of SOC is primarily on the Os ion,
and changes the magnetic character considerably in
\scoo~ but little in \scro.
}
\begin{center}
\begin{tabular}{ccccccccccccc}\hline\hline
      &\multicolumn{3}{c}{LSDA} & &\multicolumn{7}{c}{LSDA+SOC}&
\\\cline{2-4}\cline{6-13}
      &\multicolumn{4}{c}{}&\multicolumn{3}{c}{Spin}& 
      &\multicolumn{3}{c}{Orbital} & \\\cline{6-8}\cline{10-12}
Compound &Cr &Os/Ru& $M_{S}$& &Cr &Os/Ru& $M_{S}$ & &Cr & Os/Ru & $M_{L}$ 
     & $M_{net}$ \\\hline
\scoo &2.22&$-1.58$&0.00& &2.194 & $-1.312$ &0.399 & 
      & $-0.025$ & 0.173 & 0.142 & 0.541\\
\scro &2.08&$-1.48$&0.00& &2.079 & $-1.474$ &0.025 & 
      & $-0.048$ & 0.031 & $-0.020$&$-0.005$\\
\hline\hline
\end{tabular}
\end{center}
\label{table1}
\end{table*}
\begin{figure}[tbp]
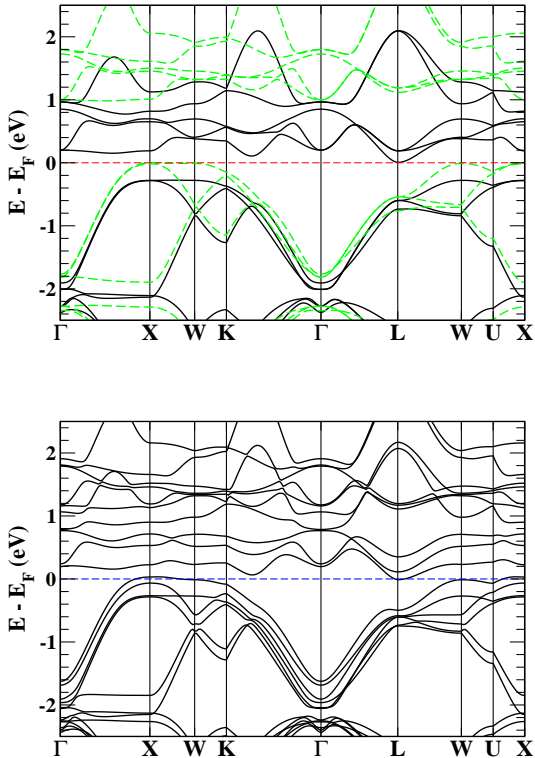

\vskip 6mm
\resizebox{7cm}{4.5cm}{\includegraphics{Fig1a.eps}}
\vskip 10mm
\resizebox{7cm}{4.5cm}{\includegraphics{Fig1b.eps}}
\caption{(color online) Top panel: LSDA band structure 
Sr$_2$CrOsO$_6$, excluding spin-orbit coupling,
in the distorted $R\bar{3}$ structure for each spin channel
(dashed and solid lines for the majority and minority channels,
respectively).
This plot contains the Cr $d$ bands and Os $t_{2g}$ manifold.
The O $p$ bands lie in the range $-8.5$ to $-2$ eV, and
the Os $e_g$ manifold lies above 4 eV due to large crystal field
splitting.
Note the flat bands along the $X-W$ lines near $E_F$,
which lead to a quasi-two-dimensional DOS at either
spin channel (see Fig. \ref{osdos}).
The symmetry labels follow the conventional fcc notation.
The horizontal dashed line indicates the Fermi energy $E_F$ set to zero.
Bottom panel: The band structure of Sr$_2$CrOsO$_6$ 
with spin-orbit coupling included.  The slightly
negative gap is due to small overlap of the conduction at L with the
upper valence band at X.  The optical gap is $\sim$0.25 eV.
 }
\label{osband}
\end{figure}
\begin{figure}[tbp]
\rotatebox{-90}{\resizebox{6.5cm}{7.5cm}{\includegraphics{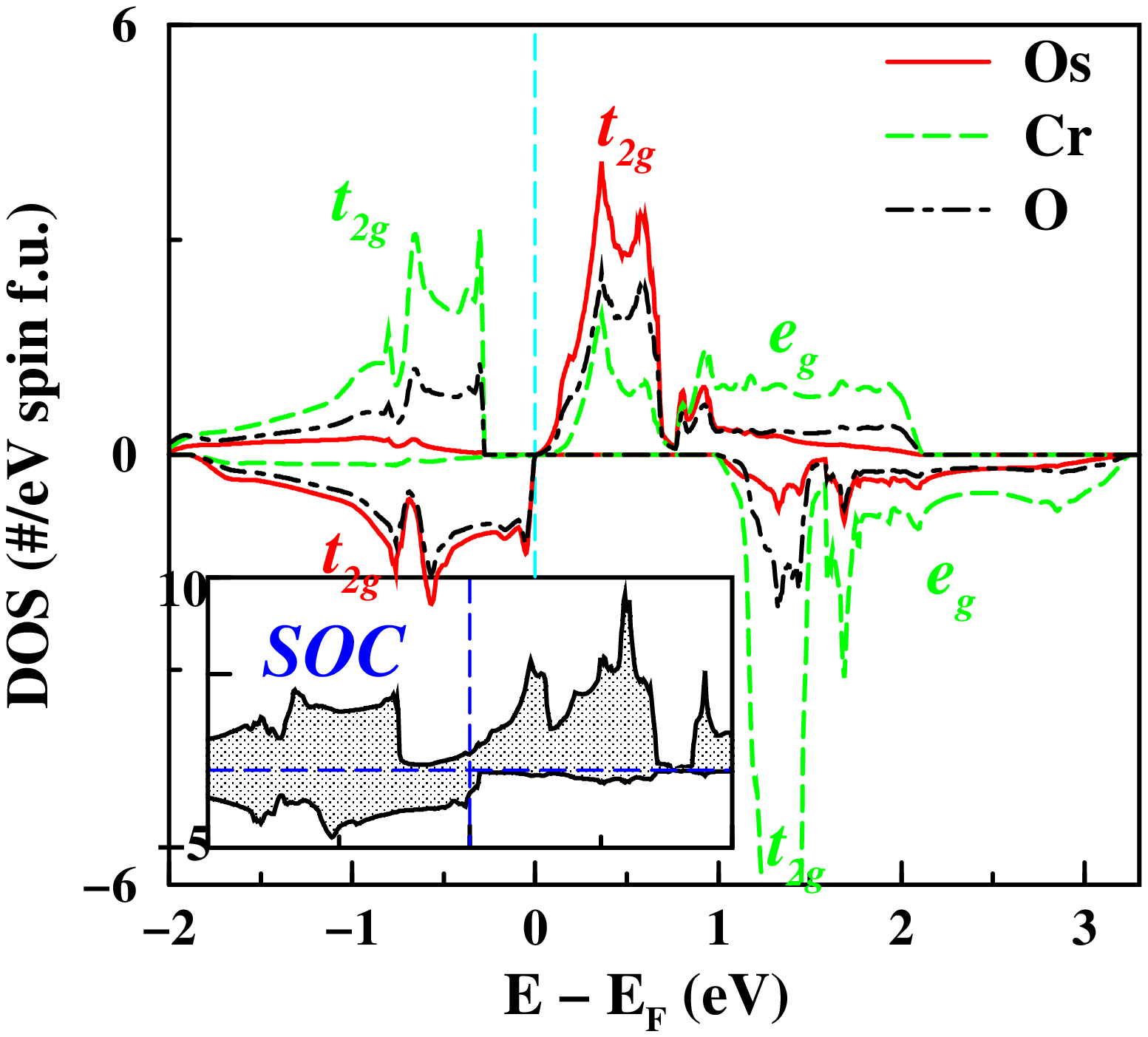}}}
\rotatebox{-90}{\resizebox{6.5cm}{7.5cm}{\includegraphics{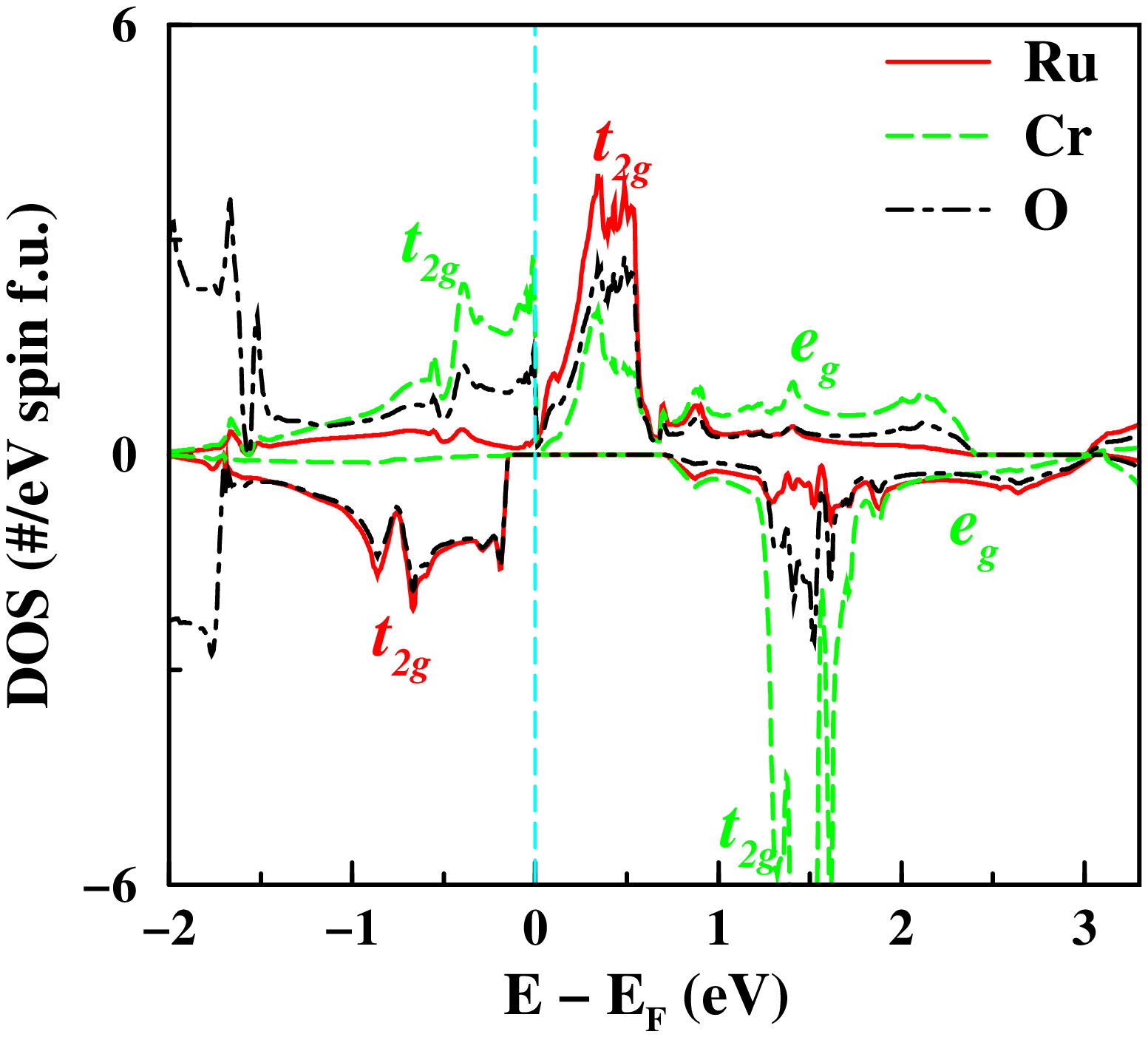}}}
\caption{(color online) Top panel: LSDA (without SOC) 
 atom-projected densities of states (DOS) per a formula unit 
 for each spin channel of Sr$_2$CrOsO$_6$ in the distorted structure.
 The occupied Os $t_{2g}$ manifold has 50\% O character in the whole
 range, while in the occupied Cr $t_{2g}$ manifold O contributes 
 less than half of Cr contribution to DOS.
 The vertical dashed line denotes $E_F$.
 Inset: LSDA+SOC total DOS (in the same units as above) 
 in the range of 1 eV above and
 below $E_F$, where SOC effects are dominant, 
 indicated by the vertical dashed line.
 The horizontal dashed line represents zero DOS.
Bottom panel: similar plot for Sr$_2$CrRuO$_6$ in the distorted
 structure.
 The O $p$ bands locate at the range $-7.5$ eV to $-1.5$ eV
 (not shown here).
 Unlike for SCOO, the DOS of SCRO is hardly affected by SOC.
}
\label{osdos}
\end{figure}
The LSDA band structure near E$_F$ (without SOC) of SCOO, which has antialigned
$S=\frac{3}{2}$ moments on both Cr and Os and hence zero net moment,
is shown in Fig. \ref{osband} and agrees with that of Krockenberger 
{\it et al.}\cite{krock}
The fully occupied Os (`down', by our choice) and Cr (up) majority 
$t_{2g}$ manifolds each
have identical width of 1.8 eV, shifted by 0.3 eV in energy.
The lowest spin-up conduction band at L is degenerate with the
filled Os $t_{2g}$ band maximum (flat along X-W), defining E$_F$ at the point
of (accidentally occurring) zero gap.
The larger transition metal--O hybridization for a $5d$ ion
compared to that of a $3d$ ion reduces the Os spin moment
from 3 $\mu_B$ by almost a factor of two, even though both Cr and Os ions have
occupied $t_{2g}$ states corresponding to $S=3/2$.

Now we will focus on results of the SCOO zero temperature R$\bar{3}$ structure.
As shown in Table \ref{table1}, the  Cr spin moment of 2.2 $\mu_B$
is completely compensated with the Os spin moment of $-1.6$ $\mu_B$
and the six O moments. 
As displayed in the corresponding DOS given in the top panel of 
Fig. \ref{osdos},
each spin channel separately has a gap, 0.4 eV for the spin-up and 1 eV 
for spin-down, between the Cr and Os $t_{2g}$ manifolds.
The majority and minority states show different
hybridization, resulting in differing crystal (ligand) field and
exchange splittings on Cr and Os.  
The crystal field splittings $\Delta_{cf}$ are (roughly): 
Cr, majority 2.5 eV, minority 1 eV; Os: 2.5-3 eV for both directions.
The exchange splitting 
in the Os ion (just over 1 eV) is half of that of Cr ion.

The distinguishing feature of SCOO is that the Os $t_{2g}$ bandwidth
is equal to its exchange splitting, such that the zero gap lies
`between' the corresponding up and down bands.
Upon including SOC,
the  Cr moment (with its small SOC) is almost unchanged.  The effect on Os
is substantial, however; due to mixing with states of the 
opposite spin, the spin moment is reduced by 0.27 $\mu_B$
and an orbital moment of -0.17 $\mu_B$ is induced.  The total net moment,
0.54 $\mu_B$ reflecting both spin and orbital compensation, is reasonable
close to the observed value\cite{krock} of 0.75 $\mu_B$.  (The unusual
temperature dependence indicates non-standard behavior of the magnetism
in SCOO, so it may be premature  to expect close agreement.)
The total DOS, given in the inset of the top panel in Fig. \ref{osdos}, 
indicates filling of the up-spin gap due to Os spin-mixing.

Krockenberger {\it et al.}\cite{krock} have reported moments of \scoo~obtained
from neutron scattering, which are net (spin+orbital) values for
Os and Cr.  These moments should not be expected to be identical to the
calculated values, both because the theoretical specification involves
and assignment of the spin density using inscribed atomic spheres (and some
spin polarization in the interstitial region not assigned to either), and 
because the neutron fitting takes no account of the difference of spin
and orbital form factors.  The experimental values are $M = 2.03~(-0.73)
\mu_B$ for Cr (Os).  Our calculated values are $M = 2.17~(-1.14) \mu_B$,
respectively, which are close to those calculated earlier.\cite{krock}  
The difference for the Os ion (which amounts to about
+25\% and -25\% from the mean of -0.95$\mu_B$) is large enough to be
bothersome, and should be the object of further study.  It has recently
been observed that the net moment on Os$^{7+}$ in Ba$_2$NaOsO$_6$ is
affected very peculiarly\cite{bnoo} 
by the large SOC on Os, and although the Os$^{5+}$
ion will differ substantially, the orbital component will need to be
taken into account explicitly to clarify this discrepancy.

\subsection{The Ru analog Sr$_2$CrRuO$_6$}
SCRO shows some distinctions from SCOO, small but sufficient to change the
character of the spectrum.  Both Cr and Ru moments are 7-9\% smaller
than in SCOO, but the system is still representative of compensating
$S=\frac{3}{2}$ moments with vanishing net spin. Relative to the Ru
(minority) bandwidth, the Ru exchange splitting is larger, and the 
on-site energy on Ru is enough lower to alter the distinguishing 
features of the states at low energy.
The result is that E$_F$ is pinned
in a `zero gap' between the occupied Cr $t_{2g}$ bands and 
unoccupied Ru $t_{2g}$ bands
(there is a actually a slight $\sim 0.1$ eV band overlap).

A crucial difference from SCOO is that the zero total moment of SCRO (due to
half metallicity)
survives SOC:
the slight net spin moment of 0.025 $\mu_B$ is
compensated by the induced orbital orbital moment of $-0.020$ $\mu_B$,
see Table \ref{table1}. 
This size of moment is negligible compared to other uncertainties, 
and we conclude
that SCRO remains effectively half metallic in
the presence of SOC.

\subsection{Metal versus insulator question}
To address the reported insulating character in SCOO (versus the low
DOS semimetal result obtained above), 
we applied intra-atomic Coulomb repulsion $U$ on Os (or Ru in SCRO)
ion using LDA+U method\cite{amf,fll} implemented in FPLO.\cite{fploU}
As expected beginning from the zero gap in SCOO, 
for $U$ as small as 0.5 eV an insulating phase is obtained.
Across the metal-insulator transition and into the insulating phase,
the compensated moment in both SCOO and SCRO
remains unchanged for all $U$.  Here the opening of the gap is not
such as to produce a Mott insulator (it is a band insulator), but
rather in the mold of a `energy gap correction' to LSDA, which might better be
accomplished by self-energy corrections suitable for itinerant state
(if it is needed at all).

An important implication is that the Os $5d$ electrons in SCOO are not in
localized, strongly correlated states but closer to itinerant.  If this
is the case, pressure will more readily produce metalization 
and lead to
a CHM (degraded by SOC).  Calculations confirm that the main effects of pressure
are to increase bandwidths, thereby reducing the moments and the associated
exchange splittings and generally increasing itineracy. The differences 
displayed by SCRO relative to SCOO suggest that
it may not be insulating even at ambient pressure, in which it provides
the first example of a CHM, not even blemished noticeably by SOC. 
The strong hybridization of the Os and Ru $d$ states, which is the cause
of itineracy, is evident in the moment that opposes that of Cr: it is
$\sim$1.4 $\mu_B$ on the metal, and 0.7 $\mu_B$ distributed over the neighboring
O ions (Table \ref{table1}).   This strong ${\cal T}$-$d$--O $2p$ mixing,
together with the robust S=$\frac{3}{2}$ moments,
is almost certainly at the root of the remarkably high magnetic ordering
temperature of 750 K. 

\subsection{Fixed Spin Moment Studies}
\begin{figure}[tbp]
\rotatebox{-90}{\resizebox{6.5cm}{7.5cm}{\includegraphics{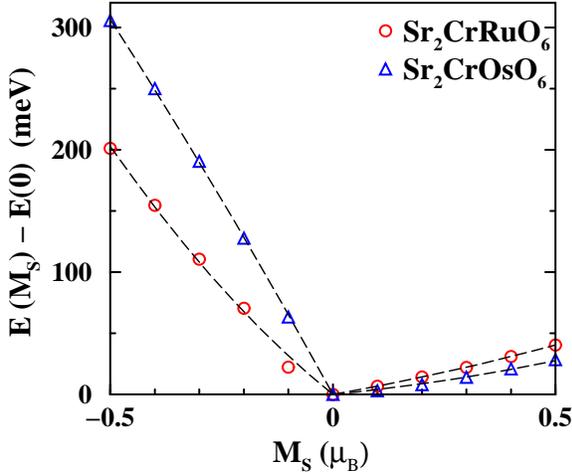}}}
\caption{(color online) LSDA fixed spin moment $M_S$ calculations.
 Since the minority channels have much larger gap in both systems,
 the plots are highly asymmetry with respect to the ground state
 occurring at $M_S=0$.
 The dashed lines indicate fitting curves using a formula
$E(M_S)-E(0)=aM + bM^2$.
 }
\label{fsm}
\end{figure}
In a conventional, unpolarized metal fixed spin moment (FSM) 
calculations\cite{oka,krasko}
provide at small M the (interaction enhanced) susceptibility $\delta E(M)=
\frac{1}{2}\chi^{-1}M^2 + ...$, and gives the Stoner
instability if the energy {\it decreases} with moment $M$.
At larger M stable or metastable magnetic states may be uncovered. 
The behavior is quite different in a half metal, where the spin 
densities (and the ground state itself)
are invariant to any applied magnetic
field that does not cause the Fermi level in the metallic channel 
to cross a band edge of the insulating channel.\cite{eschrig}
Once the crossing occurs, one can again carry out FSM calculations
to analyze the $E(M)$ relation and the differential susceptibility.
While $E(M)$ is analytic (and even in $M$) in a nonmagnetic metal,
in a HM $E(M)$ is non-analytic at $M$=0 and is asymmetric, and may be
strongly asymmetric.  The point single $M$=0 (in SCOO, since it is
spin compensated) corresponds to a huge range
of magnetic field $\Delta B = E_{gap}/\mu_B$.

We choose the Cr moment direction to define `up' in the following discussion, 
hence a positive field is parallel
to the Cr moment.  
At zero field the net moment is zero, and positive
and negative fields do not change the moment until they become large 
enough to shift the Fermi level across a `down' spin band edge. 
At the onset of change in the moment the behavior in a HM differs 
from that of conventional FSM behavior 
because of the 
strong density of states variation 
\begin{eqnarray}
N(E)\propto m^* \sqrt{|E-E_o|} 
\end{eqnarray}
at the band edge, where $m^*$ is the effective mass. 
As a result the differential susceptibility
$-d^2{\cal E}/d^2H$
at the edge is non-analytic in M.
In SCOO, however, the top of the occupied Cr $t_{2g}$ bands display 
essentially two-dimensional step-function
behavior shown in Fig.~\ref{osdos}, top panel, 
(not uncommon in perovskite structures), and beyond the 
onset the DOS varies slowly.
The form of the differential susceptibility at the conduction band edge 
(positive moment) is 
\begin{eqnarray}
\chi(M) = -d^2{\cal E}/d^2H \propto \alpha + \gamma m^* \sqrt{M},
\end{eqnarray}  
where $\alpha, \gamma$ are material-dependent constants.
The calculated
FSM $E(M)$ curves for both compounds are displayed in Fig. \ref{fsm}.  
The non-analytic
behavior is confined to low moment ($M\approx$0) and is not clear on this scale.  

The strong
asymmetry is striking. For positive applied field, the up spin states in
Fig.~\ref{osdos} move down with respect to the down states.  Because the 
exchange splitting is much smaller on Os than on Cr, the moment on Os
decreases (transfer of electrons from Os $t_{2g}$ down to Os $t_{2g}$ up)
and the energy cost is relatively small, see Fig. 3 for positive $M_s$. 
For negative applied field, the relative shift of states is in the opposite
direction, and the shift in filled states is from Cr $t_{2g}$ up to Cr
$t_{2g}$ down, and this field direction decreases the Cr moment.  
The energy cost of
changing the Cr moment is roughly an order of magnitude larger than for Os:
the Os moment is much `softer' 
than is that of Cr.  
Interestingly, the Cr moment
softens by one-third in SCRO, due to differences in hybridization, and the
Ru moment is somewhat stiffer than the Os moment. 

\section{Discussion}
We have applied first principles, full potential, all-electron methods
to assess the electronic and magnetic structures of \scoo~and \scro.
We have shown that the observed saturation moment of ferrimagnetic
\scoo~is entirely due to spin-orbit coupling, and without this coupling
\scoo~would be spin-compensated (zero net moment).  We have also studied
isostructural, isovalent \scro, which has not yet been synthesized in
ordered form.  Due to the much smaller spin-orbit coupling in Ru, this 
compound remains a spin-compensated ferrimagnetic (to within 0.02 $\mu_B$),
and therefore is a realization of a half (semi)metallic antiferromagnet.   

We have also investigated the other isovalent partner 
Sr$_2$CrFeO$_6$ using the same structures as for SCOO.
The smaller hybridization of Fe increases 
the spin moment to 3 $\mu_B$, and 
the Cr  moment decreases (1.3 $\mu_B$).
The result is a simple metallic ferrimagnet with a net spin moment of
$2.25$ $\mu_B$.

Williams {\it et al.} reported Cr-Ru disorder\cite{williams} in their
study of the SrCr$_{1-x}$Ru$_x$O$_3$ system.
However, the negligible difference of 0.01 \AA~ in the Ru$^{5+}$ and
Os$^{5+}$ ionic radii\cite{chemweb} and the identical charge state suggest
that ordered SCRO would result from the high temperature technique 
that produces (almost
perfectly) ordered SCOO.\cite{krock}  Another approach would proceed by
alternating Cr and Ru deposition in
layer-by-layer epitaxy, either 
molecular beam epitaxy or pulsed laser deposition, a process which 
produces the double perovskite structure
for $<111>$ growth.  Thus prospects should be good for
checking for the CHM state (``half metallic antiferromagnetism'') in \scro.  

Finally, we have performed the first (to our knowledge) extension of the
fixed spin moment technique to a half metallic compound.  We describe how
this technique provides different and still important information for a
half metal.  For \scoo, it indicates that longitudinal fluctuation in the
magnetic moment of Os in \scoo~is about ten times as large as for the Cr
moment.  In \scro, this dissimilarity is reduced to about a factor of five.

\section{Acknowledgments}
We acknowledge clarification of the crystal structure by
M. Reehuis and M. Jansen, communication with W. H. Butler on
fixed spin moment calculations in half metals, and discussions with C.
Felser on these compounds.  This work was supported
by DOE grant DE-FG03-01ER45876, and interaction within DOE's
Computational Materials Science Network is acknowledged.

\end{document}